\documentclass[a4paper]{aipproc}
\layoutstyle{6x9}
\begin{document}
\title{Soft Gluons
and  the Energy Dependence of
Total  Cross-Sections}

\author{R. M. Godbole}
{address={Centre for Theoretical Studies, 
Indian Institute of Science, Bangalore, 560 012, India},
email={rohini@cts.iisc.ernet.in},}
\author{A. Grau}
{address={Centro Andaluz de F\'\i sica de Part\'\i culas Elementales and
  Departamento de F\'\i sica Te\'orica y del Cosmos, Universidad de
   Granada, Spain},
email= {igrau@ugr.es},}
\author{G. Pancheri}
{address=
{INFN Frascati National Laboratories, Via E. Fermi 40, I00044 Frascati,
Italy},
email={pancheri@lnf.infn.it},}
\author{Y. N.  Srivastava}
{address={INFN, Physics Department, University of Perugia, Perugia, Italy},
email={yogendra.srivastava@pg.infn.it},}
\setcounter{page}{0}
\thispagestyle{empty}
\begin{flushright}                               
                                                   hep-ph/0205196  \\
                                                    IISc/CTS/16-01\\ 
                                                    UG-FT-136/01 \\ 
\end{flushright}

\vskip 25pt

\begin{center}

{\bf  Soft Gluons and  the Energy Dependence of Total  
          Cross-Sections\footnote{Talk given by G.P. at QCD@work, International  Workshop on QCD, Martina Franca, Italy, June 16-20,2001}}
\vskip 25pt
  R.M. Godbole$^1$, A. Grau$^2$, G. Pancheri$^3$ and Y. Srivastava$^4$ 

\bigskip
1. Centre for Theoretical Studies, Indian Institute of Science, 
Bangalore, 560012, India. E-mail: rohini@cts.iisc.ernet.in\\
\vskip 25 pt
2.   Centro Andaluz de F\'\i sica de Part\'\i culas Elementales and
   Departamento de F\'\i sica Te\'orica y del Cosmos, Universidad de
   Granada, Spain.  E-mail igrau@ugr.es\\
\vskip 25 pt
3. INFN Frascati National Laboratories, Via E. Fermi 40, I00044 
Frascati, Italy. E-mail:pancheri@lnf.infn.it\\
\vskip 25pt
4. INFN, Physics Department, University of Perugia, Perugia, Italy.\\
E-mail: yogendra.srivastava@pg.infn.it\\
\bigskip
           Abstract
\end{center}

\begin{quotation}
\noindent
We discuss the high energy behaviour of total
cross-sections for protons and photons, in a QCD based framework with
particular emphasis on the role played by soft gluons.
\end{quotation}

\newpage

\begin{abstract}
We discuss the high energy behaviour of total
cross-sections for protons and photons, in a QCD based framework with
particular emphasis on the role played by soft gluons.
\end{abstract}
\maketitle

\section{Introduction}
Energy dependence of  hadronic total cross-sections has fascinated 
 particle physicists for decades now.  In this talk we address 
a number of questions which arise when studying total hadronic 
cross-sections, namely
\begin{itemize}
\item Is it possible to study the energy dependence of the
cross-sections for 
$pp$, $p \bar p$, $\gamma p$ and $\gamma \gamma \rightarrow { hadrons}$
in the same phenomenological/theoretical framework?
\item What governs  the energy dependence of these total cross-sections?
\item What is the  role played by the electromagnetic  form factors in
 the description of the total cross-section?
\end{itemize}

\begin{figure}[ht]
\includegraphics[scale=0.7]{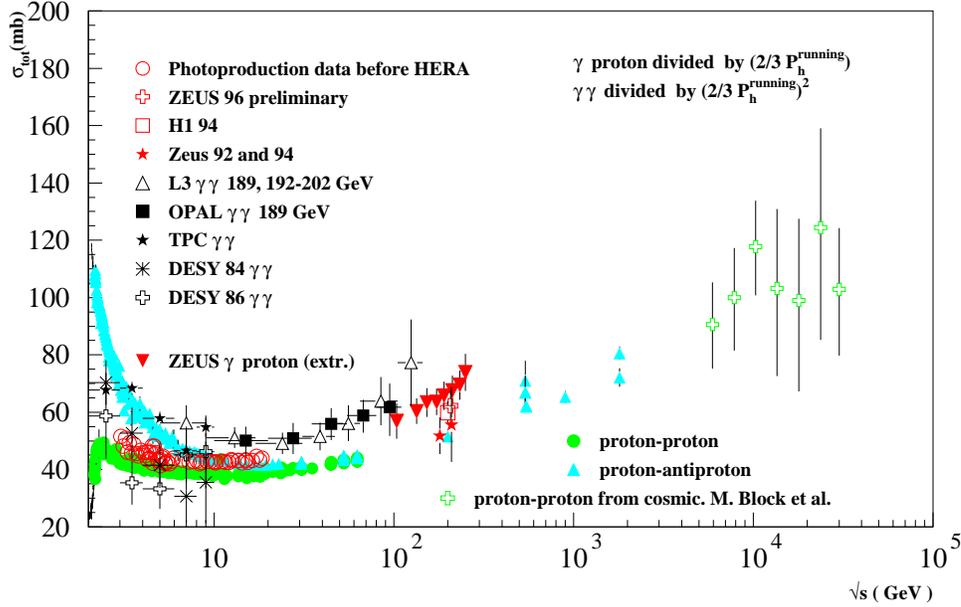}
\caption{A compilation of $pp$,$p\bar{p}$,$\gamma p$
and $\gamma \gamma$ total cross sections with scaling 
factors described in the text.}
\label{all}
\end{figure}

The first question about treating {\it together}  the $pp, p \bar p$ 
case on the one hand and the $\gamma p, \gamma \gamma$ case on the 
other,  arises naturally as the `hadronic' structure~\cite{review} 
of the photon has now been established in both $e^+e^-$ and $ep$ experiments 
conclusively~\cite{expts}. 
Further, the photonic partons seem to have
nontrivial effects on the photon-induced processes at high 
energies~\cite{Drees-Godbole}. Equally importantly,  along with the data
already available for the $pp,p\bar p$
case~\cite{CDF-D0},
data have become available on 
total cross-sections for photon-induced processes  reaching up to high
$\gamma$ energies, $\gamma p$ and $\gamma \gamma$ processes being 
studied in $ep$\cite{H1,ZEUS,DIS,ZEUS-prelim},  
and $e^+e^-$~\cite{L3,OPAL} collisions respectively.
In Fig.\ref{all} we show a compilation of these 
proton and photon total cross sections, including cosmic ray data as 
well~\cite{martincosmic}.
In order to put all the data on the same scale\cite{ourplb,epjc}, we
have used a multiplication 
 factor suggested by quark counting and Vector Meson
Dominance\cite{FLETCHER}, namely
a factor 
$2/3 \sum_{V=\rho,\omega,\pi} \left(4\pi \alpha_{QED}/f^2_V \right)$.
Using a running $\alpha_{QED}$, the VMD factor ranges from $1/250$ at low
energy to $1/240$ at HERA energy. Square of this factor enters the
photon-photon cross-sections.


At first glance,  these data  raise two questions:
(i) whether the   $\gamma \gamma $ total cross-section rises faster
than the others and (ii) whether these various sets of data are
mutually consistent (at least at low energies) with the factorization
hypothesis \cite{aspen}. The uncertainty in the
normalization of photon processes does not yet allow for a definite
answer, but the photon-photon cross-sections do seem to
be rather different, both from the point of view of the
normalization~\cite{aspen} as well as the rise \cite{epjc,lundus,albert}.

The next question is whether and how can we understand  these data with
our present means to deal with QCD. It appears that not all but many
of the observed features are quantitatively obtainable from QCD.
Our present goal is to obtain a QCD description of the initial decrease
and the final increase of total cross-sections through soft gluon
summation (via Bloch-Nordsieck Model) and mini-jets. Thus, our
physical picture includes multiple parton collisions and soft gluons
dressing each collision. We shall describe in the following sections
details of the theoretical model proposed.

\section{A QCD Approach}
\begin{figure}[ht]
\includegraphics[scale=0.5]{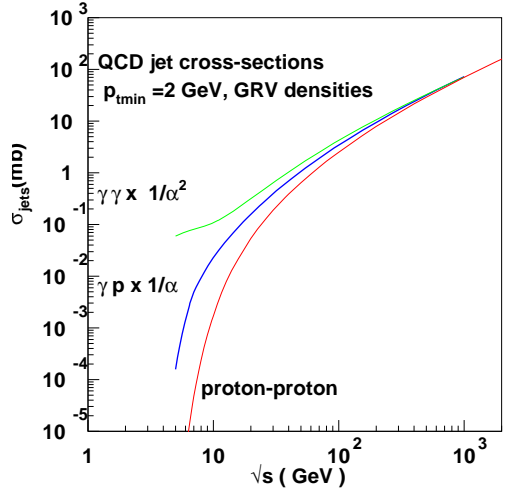}
\caption{Minijets: Integrated jet cross-sections}
\label{minijets}
\end{figure}

The task of describing the energy
behaviour of total cross-sections can be broken down into three parts:
\begin{itemize} \item the rise
\item the initial decrease
\item the normalization
\end{itemize}
The rise~\cite{therise}
can be obtained using the QCD calculable
contribution from the parton-parton cross-section, whose total
yield increases with energy, as shown in Fig.(\ref{minijets}),
where the jet cross-sections for proton-proton, $\gamma\ p$ 
and $\gamma \gamma$
are scaled by a common
factor $\alpha$.

In all cases, in particular for the proton case (where there are no direct
scattering terms),  one observes that $\sigma_{jet}$ rises too fast for
the observed values of $\sigma_{tot}$ (less than 100 mb at the Tevatron)
and that other terms, due to soft interactions, are missing. For a
unitary description, the jet cross-sections are embedded into the eikonal
formalism \cite{eikminijets}, namely one writes
\begin{equation}
 \label{eiktot}
\sigma^{\rm tot}_{pp(\bar p)}=2\int d^2{\vec b}
[1-e^{-\chi_I(b,s)}cos(\chi_R)]
\end{equation}
where the eikonal function $\chi=\chi_R+i\chi_I$ contains both the
energy and the  transverse momentum dependence of matter distribution
in the colliding particles, through the  impact parameter distribution in
b-space\cite{bn}. The simplest formulation with  minijets to drive the  rise,
in conjunction with eikonalization to ensure unitarity,  is:
\begin{equation}
2 \chi_I(b,s)\equiv n(b,s)=A(b) [ \sigma_{soft}+\sigma_{jet}]
\end{equation}

The  normalization depends both upon  $\sigma_{soft}$ and the b-distribution.
A very first working hypothesis is that the impact parameter distribution 
follows the matter distribution inside hadrons, namely that it is given by the
Fourier transform of the electromagnetic form factors of the colliding
particles, i.e.
\begin{equation}
A_{ab}(b)\equiv A(b;k_a,k_b)= {{1}\over{(2\pi)^2}}
\int d^2 {\vec q} e^{iq\cdot b}
{\cal F}_a(q,k_a){\cal F}_b(q,k_b)
\end{equation}

\begin{figure}[ht]
\includegraphics*[scale=0.5]{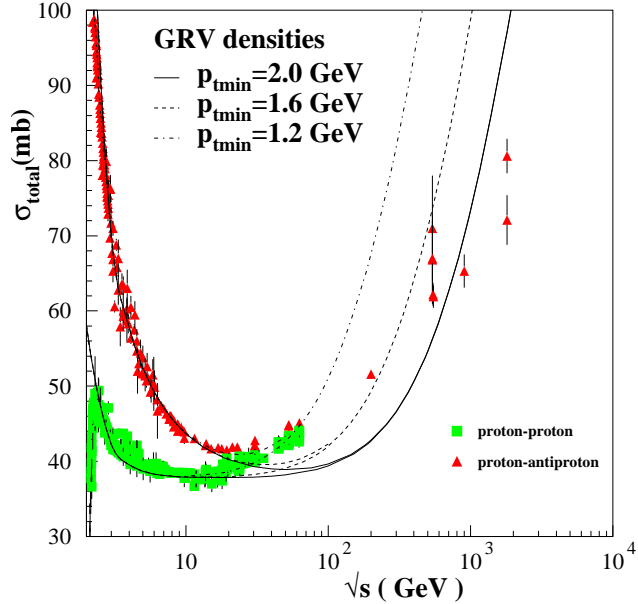}
\caption{Total cross sections for $pp$ and $p\bar{p}$
from EMM for various $p_{tmin}$.}
\label{EMMprotons}
\end{figure}

With such  hypothesis, it is possible to describe the early rise,
which takes place around $10 - 50$ GeV for proton-proton and
proton-antiproton scattering, using GRV~\cite{GRV}
densities for the protons and a
transverse momentum cut-off in the jet cross-sections,
$p_{tmin}\simeq  1 $ GeV, but then the cross-sections begin to rise too
rapidly. One needs a $p_{tmin}\approx 2$ GeV in order to reproduce the
Tevatron data, with the  drawback, however,   that  one misses the early
rise. In Fig.(\ref{EMMprotons}) we show a straightforward application of the
Eikonal Minijet Model (EMM), with different values of $p_{tmin}$, to
illustrate this feature.

A possible way to circumvent this problem lies in the use of soft
gluons instead of form factors, but before turning to the issue of how
to reproduce the early rise in proton-proton as well as the further
Tevatron data points, we discuss the question of the photon
cross-sections.

\section{Photon processes and Minijets}

Photo-production and extrapolated data from Deep Inelastic Scattering
(DIS) can be described through the same simple eikonal minijet model, 
with the relevant parton densities for the jet cross-sections,  
scaling~\cite{collins} the non perturbative part given by  $\sigma_{soft}$
with the VMD and quark counting factor discussed above. The minijet
cross-sections are then embedded into the eikonal formalism, with proper
choice of impact parameter distribution.  One needs a b-distribution of
partons in the photon, which can be chosen to be  a meson-like form
factor.

\begin{figure}[ht]
\includegraphics[scale=0.5]{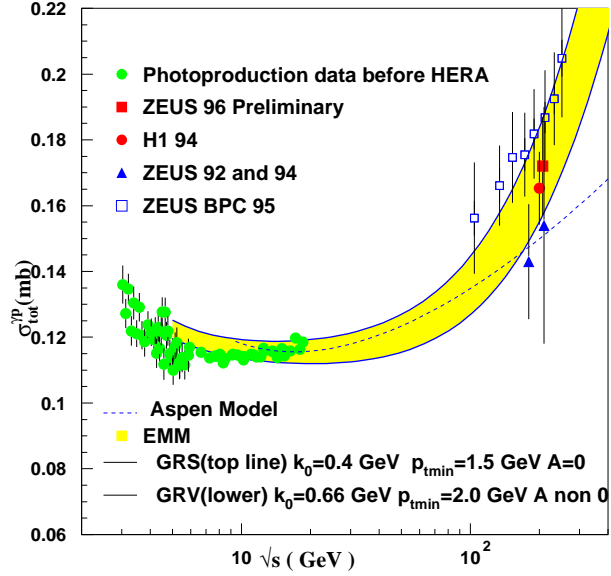}
\vspace{1cm}
\caption{Description of  photoproduction data with the EMM (band) and
Aspen model (dotted line)}
\label{hera}
\end{figure}

The result is shown in Fig.(\ref{hera}),
where the band corresponds to different sets of
model parameters, with both GRV~\cite{GRVPH} and GRS~\cite{GRS}  densities
for the photon, and the dotted line corresponds to the predictions of the
so-called Aspen Model\cite{aspen}. The low energy region
is obtained using quark counting and VMD from the proton data, while
the high energy part is obtained from the QCD minijet cross-section and
the impact parameter distribution from proton and pion-like form factors.
As discussed in \cite{ourplb}, the scale parameter $k_0$ in the
photon form factor is allowed to vary in  the range $0.4 - 0.66$ GeV.

One encounters the same problem as in proton-proton case, albeit
in a less severe form. When the parameters of the EMM are chosen so as
to reproduce the low  as well as  the high energy data, the early rise
is not well described. Modelling of $\gamma\ p$ data is further
complicated, however, by the existence of data extrapolated from DIS~\cite{DIS}
which lie above, but within 1$\sigma$,  from recent photoproduction
measurements~\cite{ZEUS-prelim}.  Using a set of parameters consistent with 
those used to obtain the band of Fig.\ref{hera}, one can now attempt a 
description of
photon-photon collisions and make predictions for future linear and photon
colliders.

\begin{figure}[ht]
\includegraphics*[scale=0.5]{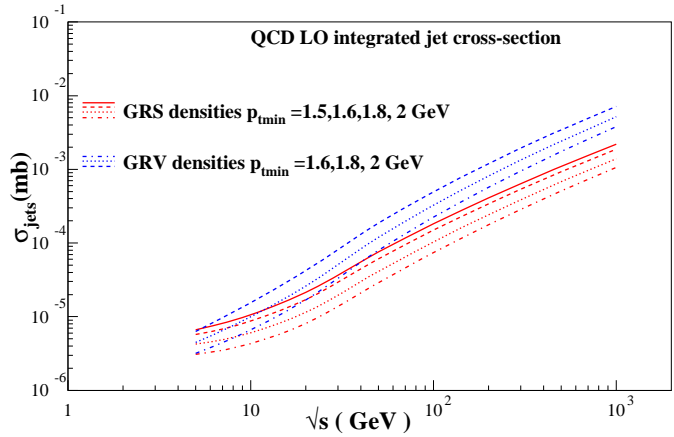}
\caption{Minijets in photon-photon collisions}
\label{minigamgam}
\end{figure}

\begin{figure}[ht]
\includegraphics[scale=0.5]{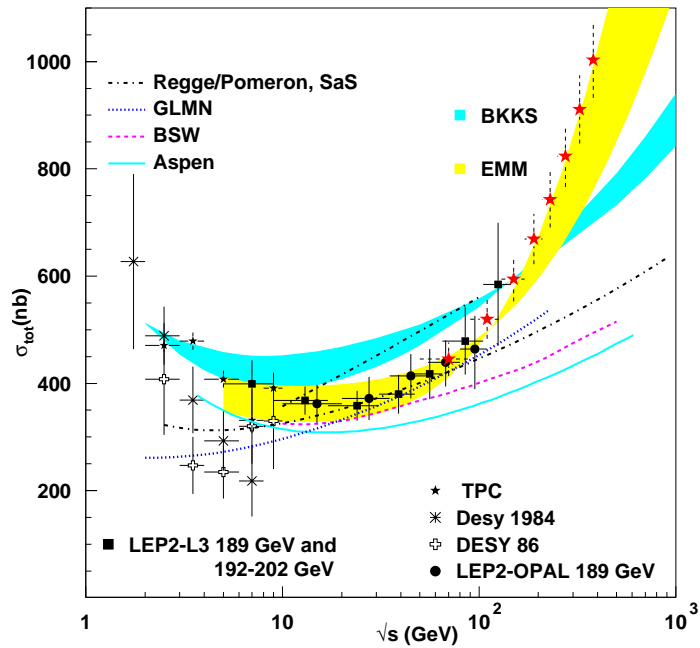}
\caption{Photon photon total cross section data compared with various
models.The stars at high photon-photon energies correspond
to pseudo-data points extrapolated \protect{\cite{LCnote}}
from EMM predictions}
\label{allmodels}
\end{figure}

As before, one starts with the mini-jet cross-sections, for various parton
densities and different values of $p_{tmin}$, as shown in 
Fig.(\ref{minigamgam}). Note that the set of curves which lie higher at 
higher energies correspond to the GRV densities.
These minijets are then embedded into the eikonal, with parameters
consistent \cite{lcws00} with the $\gamma \ p$ band shown in Fig.(\ref{hera}).
Present LEP data are shown in Fig.(\ref{allmodels})
where  EMM predictions~\cite{ourplb,epjc,lcws00} are compared with those 
from various models \cite{aspen,DL,sjostrand,maor,ttwu,
bkks} 
which have been proposed to
describe $\gamma \gamma $ total cross-sections. 
The uncertainty in the predictions of photon-photon collisions is
reflected in the uncertainty in $e^+e^-\rightarrow hadrons$, albeit, in
such case, the difference between the predictions of different models 
for  $\gamma \gamma$  total cross-section  is, at the end, 
at most a factor 2, even at TESLA energies\cite{lcws00,LCnote}.


\section{The taming of the rise through soft gluon summation}

The fast rise due to mini-jets and the increasing
number of gluon-gluon collisions as the energy increases,
can be reduced if one takes into account that soft gluons,
emitted mostly by the initial state valence quarks, give rise to  
an acollinearity between the partons which reduces the overall
parton-parton luminosity. That is, as the energy increases, the larger
phase space available for soft gluon emission implies more and
more acollinearity and thus a reduced collision probability. This is the
physical picture underlying the eikonal minijet model with Bloch-Nordsieck
resummation\cite{bn}. In this model, the impact parameter distribution of
partons is the (normalized) Fourier transform of the total transverse
momentum distribution of valence quaks, obtained through soft gluon
resummation,
i.e.
\begin{equation}
A(b,s)={{e^{-h(b,s)}}\over{\int d^2{\vec b}\ e^{-h(b,s)}}}
\label{ab}
\end{equation}
with
\begin{equation}
h(b,s)=
\int_{k_{min}}^{k_{max}} d^3 {\bar n}(k)[1-e^{-i{\vec k_\perp}\cdot {\vec b}}]
\label{hb}
\end{equation}
where $d^3 {\bar n}(k)$ is the single soft gluon differential distribution
and the integral runs, in principle, from zero to the maximum
kinematic limit. Phenomenological applications of this expression
encounter two main problems, one of theoretical origin, the other more
of a phenomenological nature, 
namely, on the one side,  a lack of our knowledge of the infrared behaviour
of $\alpha_s$ , and, on the other,  the unavailabily of 
reliable unintegrated parton
distributions, i.e. parton distributions before the integration of their
initial transverse momentum. The second difficulty can be
phenomenologically overcome by averaging
 the function $A(b,s)$ over the parton densities to obtain
the total number of collisions as
\begin{equation}
n(b,s)=A_{soft}(b)\sigma_{soft}+A_{PQCD}(b,s)\sigma_{jet}^{LO}
\end{equation}
with $ A_{soft}(b)$ as in the simpler EMM (form factors), and
$A_{PQCD}(b,s)$
given by eqs.(\ref{ab},\ref{hb}). The maximum energy
for single soft gluon emission is obtained by averaging over the
valence parton densities, i.e,
$$
M\equiv <k_{max}(s)>=\\ 
{{\sqrt{s}}
\over{2}}{{ \sum_{i,j}\int {{dx_1}\over{ x_1}}
f_{i/a}(x_1)\int {{dx_2}\over{x_2}}f_{j/b}(x_2)\sqrt{x_1x_2} \int dz (1 - z)}
\over{\sum_{i,j}\int {dx_1\over x_1}
f_{i/a}(x_1)\int {{dx_2}\over{x_2}}f_{j/b}(x_2) \int(dz)}}
$$
with $z_{min}=4p_{tmin}^2/(sx_1x_2)$. The quantity $M$ 
can be calculated as a function of $s$ for different values of $p_{tmin}$.
For  $p_{tmin}$ values between 1 and 2 GeV, it ranges between 700 MeV
and 3 GeV as  $\sqrt{s}$ goes from 20 GeV to $10$ TeV. 

\begin{figure}[ht]
\includegraphics[scale=0.4]{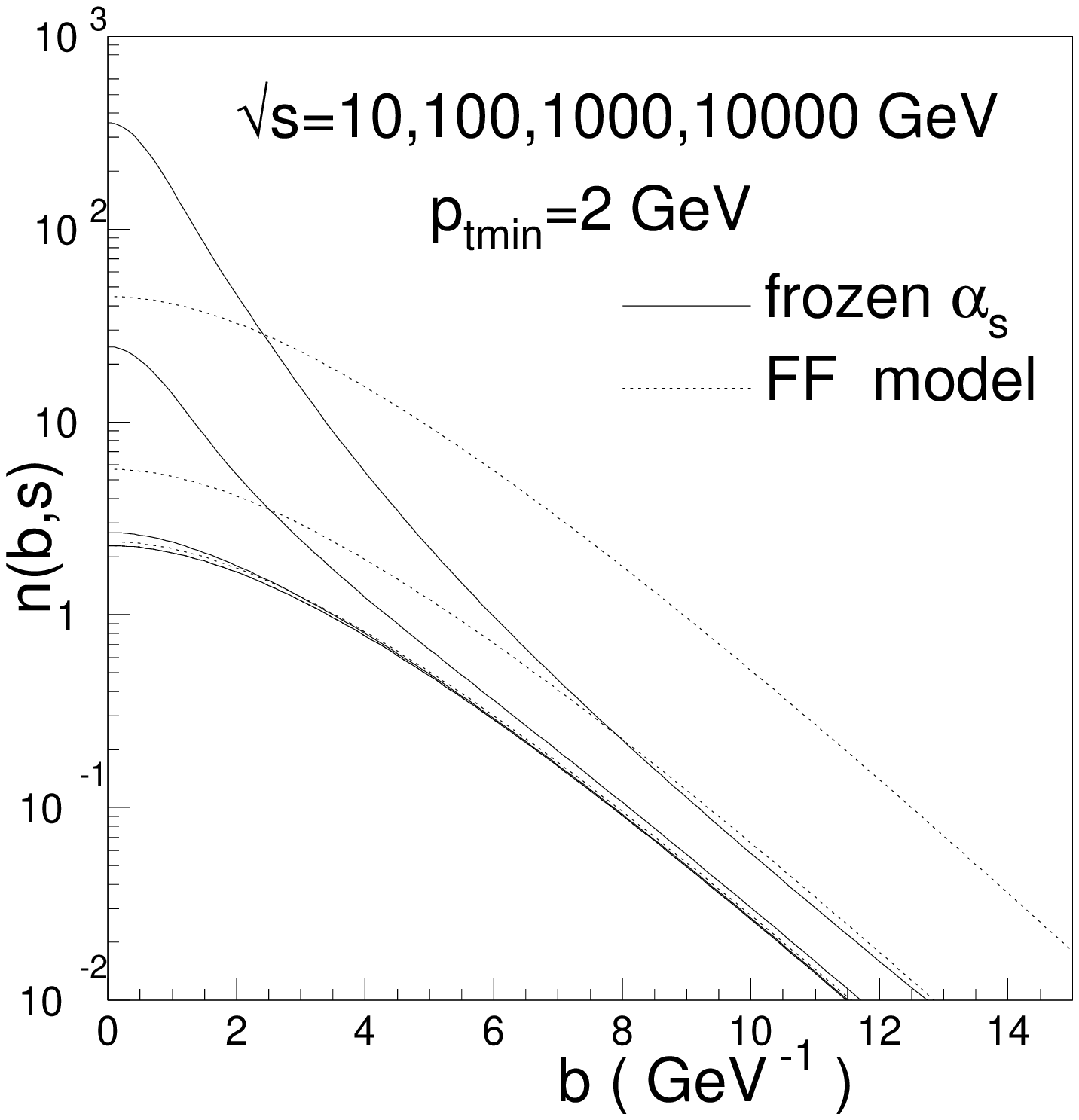}
\includegraphics[scale=0.4]{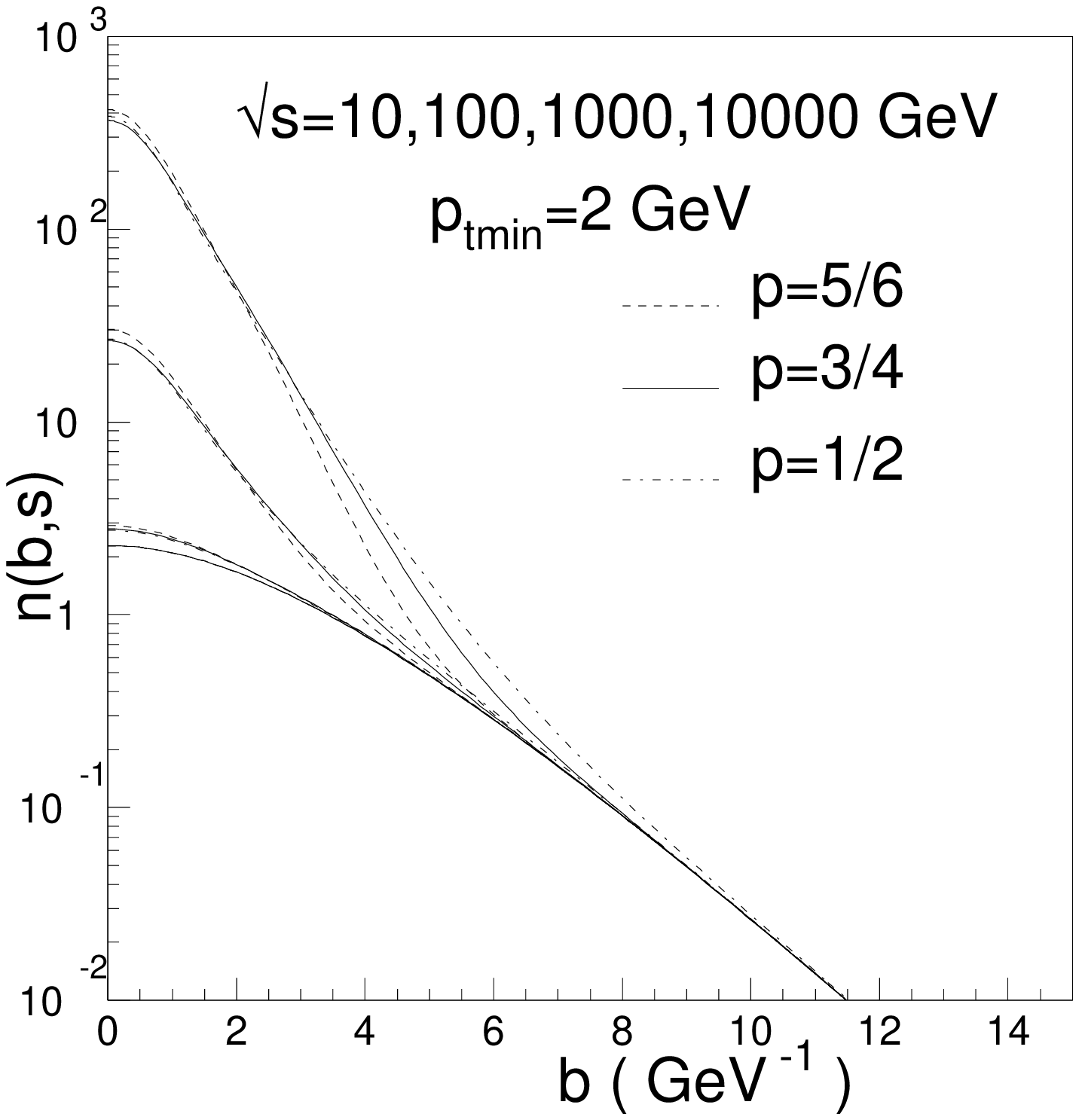}
\caption{The average number of collisions
for the frozen $\alpha_s$ case
in comparison with the form factor (FF)  model (left) 
and the singular $\alpha_s$ case (right)
for different values of the singularity parameter $p$.}
\label{nbfs}
\end{figure}

\begin{figure}[ht]
\includegraphics[scale=0.5]{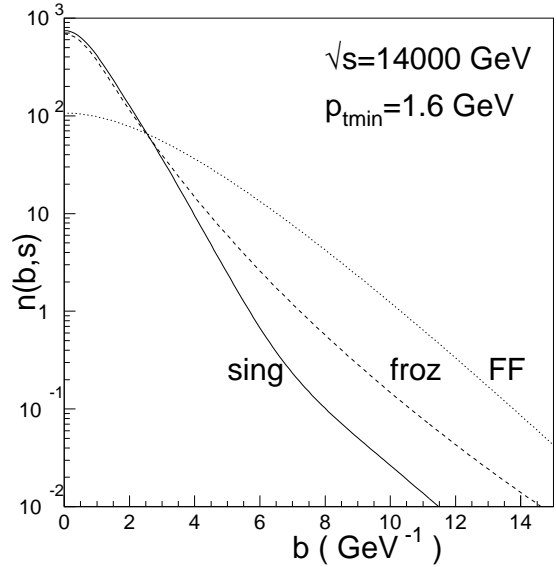}
\caption{The average number of collisions
in the form factor model and  the
Bloch Nordsieck model,
at LHC energy}
\label{nblhc}
\end{figure}

To proceed further, one also needs to specify the lower limit of
integration, or, if the value zero is assumed, the behaviour of
$\alpha_s(k_t)$ as $k_t \rightarrow 0$. Our model assumes
$k_{min}=0$ and two different trial behaviours are
utilized for the above limit, a frozen $\alpha_s$ model i.e. $\alpha_s(0)=
constant$ and a  model in which  $\alpha_s$ is singular,
but integrable\cite{yn1,yn2}.
Since a single soft gluon is never observed, one
only needs integrated quantities and, at least phenomenologically,
this model seems adequate. As discussed elsewhere \cite{bn}, the
effect of soft gluon summation is mostly to introduce an energy dependence
in the large b-behavior. In the frozen $\alpha_s$ case, the large
b-behaviour is not depressed enough, compared to the form factor case,
thus indicating the need to introduce an intrinsic transverse momentum
 cut off, namely a gaussian decrease in the b-variable. Different is
the singular $\alpha_s$ case,
where   the expression \cite{yn1}
$\alpha_s(k_\perp)={{12 \pi }\over{(33-2N_f)}}{{p}\over{\ln[1+p({{k_\perp}
\over{\Lambda}})^{2p}]}}$,
produces  an increasingly faster falloff in the
b-distribution as the energy increases.
The s-dependence of the b-distribution modifies strongly
the energy behaviour of the average number of collisions, as one can
see from Figs.(\ref{nbfs},\ref{nblhc}).

\begin{figure}[ht]
\includegraphics[scale=0.5,angle=90]{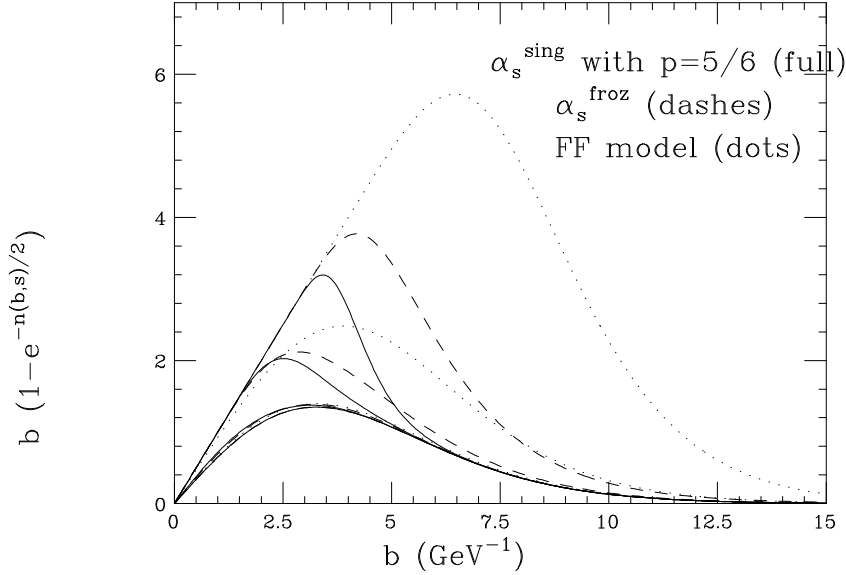}
\caption{Integrand of the eikonal function for $\sigma_{tot}$
in the three different models}
\label{int}
\end{figure}

\begin{figure}[ht]
\includegraphics[scale=0.8]{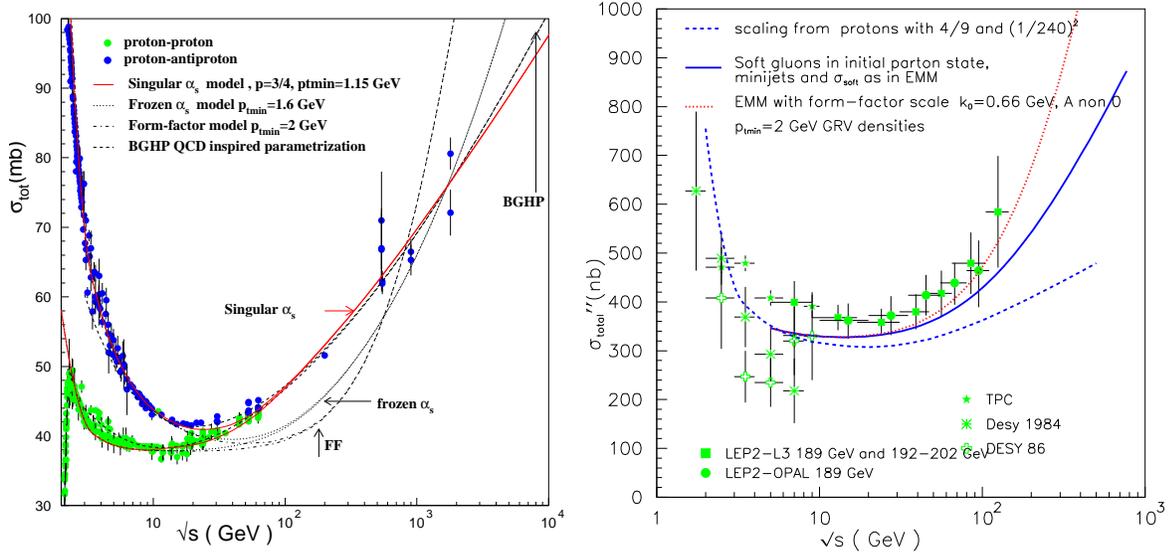}
\caption{Total cross sections for $pp$ and $p\bar{p}$ with
frozen and singular $\alpha_s$ and form factor (FF) models (left panel).
LEP data are compared with EMM with and without
soft gluons and with a curve scaled from protons (right panel). }
\label{sigtotbn}
\end{figure}

As the energy increases, the average number of collisions, relative to the
form factor model, is strongly depressed at large b, thus smaller b-values 
contribute to the total cross-section, and the cross-section remains
in general smaller than in the form factor case. In Fig.(\ref{int}), 
we show how the integrand of eq.(\ref{eiktot}) behaves as a function of
b, for $\sqrt{s} = 100, 1000 $ and $10,000$, in the three models examined 
here. Note that we take $\cos\chi_R = 1$. The peak position shifts with 
increasing energy to higher $b$ values and the area under the curve rises.
The integrand  is peaked at different $b$-values as the energy increases,
but also as the model for $A(b)$ changes. The rise with energy of the area
under the curve, i.e. the cross-section, at the same energy, shrinks for
the more singular $\alpha_s$ case. All the above features 
are illustrated in the plots given in the left panel of Fig. (\ref{sigtotbn}). 
We see that the effect of soft gluon summation in the singular $\alpha_s$ model 
reproduces quite well the early rise and the asymptotic softening. In
comparison, the frozen $\alpha_s$ model appears almost as bad as the
form factor model. The Bloch-Nordsieck model is practically
indistinguishable from more conventional curves obtained through
the Regge-Pomeron exchange \cite{DL} or the QCD inspired Aspen
model\cite{aspen}, labelled BGHP in Fig.(\ref{sigtotbn}).


The analysis of proton collisions implies that
straightforward applications of the minijet model through form factors
are unable to describe correctly the large energy rise of total
cross-sections.  On the other hand, we have seen that the EMM can
reproduce well the rise observed in $\gamma \gamma$ collisions
in the present energy range, $\sqrt{s_{\gamma
\gamma}} \approx 50-100$  GeV. But is the trend predicted by the EMM
for photon-photon scattering correct at larger c.m. energies? 
It is quite possible that  photon-photon
data are only showing the  early rapid rise, and that the rise  at higher
energies needs further corrections of the type we have described. Soft
gluons are probably necessary  in order to extrapolate to the
higher energies of future electron-positron colliders such as TESLA, CLIC,
NLC or Photon Colliders. An application of the Bloch-Nordsieck method to
the case of photon-photon collisions is shown in the right panel of 
Fig.(\ref{sigtotbn}).


\section{Conclusions}

We have described a unified approach to the calculation of total
cross-sections for protons and photons. In all cases, the driving
cause for the rise of  total cross-sections is the
energy dependent perturbative QCD parton-parton cross-section.
For photon induced processes the model seems to describe the rise adequately.
However for all  proton processes it gives a rise which
appears way too strong. Taming of the rise can
be accomplished by an energy dependent impact parameter distribution, and
different models for the infrared behaviour of $\alpha_s$ in the
 soft gluon summation have been explored.  Our phenomenological
analysis indicates a distinct preference for a singular but integrable
$\alpha_s$ which automatically produces the desired
effect of an initial intrinsic tranverse momentum of partons
in the hadrons. The resulting physical picture is that of multiple
scattering between partons, implemented by initial state
soft gluon bremmstrahlung.

\section*{Ackowledgements}

We ackowledge partial support through  EEC Contract EURODAPHNE,
TMR0169, and from Ministerio de Ciencia y Tecnolog\'\i a project FPA2000-1558.


\begin{thebibliography}{99}
\bibitem{review} R.M. Godbole, {\it Pramana}\ {\bf 51}, 217 (1998),  
{\bf hep-ph/9807402}; M. Drees and R.M. Godbole , {\it J.Phys. G}\ {\bf
21}, 1559 (1995), {\bf hep-ph/9508221}.
\bibitem{expts}
M.~Krawczyk, A.~Zembrzuski and M.~Staszel,Phys.\ Rept.\  {\bf 345}, 265 (2001)
{\bf hep-ph/0011083}.
\bibitem{Drees-Godbole}M. Drees and  R.M. Godbole, {\it Phys. Rev. Lett.}\
{\bf 67}, 1189 (1991); R.M. Godbole, Proceedings of the Workshop on{\it
Quantum Aspects of Beam Physics, Jan. 5 1998 - Jan. 9 1998, Monterey, U.S.A.},
404-416, Edited by P. Chen, World  Scientific, 1999; {\bf hep-ph/9807379}.
\bibitem{CDF-D0} CDF Collaboration, Abe, F., et al {\it Phys. Rev. D}\  
{\bf 50}, 5550 (1994).
\bibitem{H1}  H1 Collaboration, Aid, S., et al., {\it Zeit. Phys. C}
\ {\bf 69}, 27
(1995), {\bf hep-ex/9405006}.
\bibitem{ZEUS}  ZEUS Collaboration, Derrick, M., et al., {\it Phys. Lett. B}
\ {\bf 293}, 465
(1992); Derrick, M., et al., {\it Zeit. Phys. C}\ {\bf 63}, 391 (1994).
\bibitem{DIS} Breitweg, J., et al., ZEUS collaboration,
 {\bf DESY-00-071}, {\bf hep-ex/0005018}. 
\bibitem{ZEUS-prelim} ZEUS Collaboration (C. Ginsburg et al.), Proc. 8th 
International Workshop on Deep Inelastic Scattering, April 2000, Liverpool,
\bibitem{L3}  L3 Collaboration,
Paper 519 submitted to {\it ICHEP'98}, Vancouver, July 1998;
Acciarri, M., et al., {\it Phys. Lett. B}\ {\bf 408}, 450 (1997); 
L3 Collaboration, Csilling, A., {\it Nucl.Phys.Proc.Suppl. B}\ {\bf 82}, 
239 (2000);
Acciari, M., et al,, {\bf CERN-EP/2001-012}, {\it To appear in Phys. Lett. B},
{\bf hep-ex/0102025}.
\bibitem{OPAL} OPAL Collaboration. Waeckerle, F.,
{\it Multiparticle Dynamics 1997, Nucl. Phys. Proc. Suppl.B}\ {\bf 71}, 381  
(1999) edited by G. Capon, V. Khoze, G. Pancheri and A. Sansoni;
 S\"oldner-Rembold, S., {\bf hep-ex/9810011}, To appear in  the proceedings
of the {\it ICHEP'98}, Vancouver, July 1998; Abbiendi, G., et al.,
{\it Eur.Phys.J.C}\ {\bf 14}, 199 (2000), {\bf hep-ex/9906039}.

\bibitem{martincosmic}
M.M. Block, F. Halzen and T. Stanev, {\it Phys.Rev.} {\bf D62},  077501 (2000). 
M.M. Block, F. Halzen, G. Pancheri and T. Stanev,  hep-ph/0003226, {\it 25th 
Pamir-Chacaltaya Collaboration Workshop}, Lodz, Poland, November 1999.
\bibitem{ourplb}
R. M. Godbole and G. Pancheri, {\it Phys. Lett. B} {\bf 435} 441 (1998),
{\bf hep-ph/9807236} 
\bibitem{epjc}
R.M. Godbole and G. Pancheri, {\it Eur.Phys.J.C}, {\bf19}, 129 (2001),
{\bf hep-ph/0010104}
\bibitem{FLETCHER}
R.S. Fletcher , T.K. Gaisser and F.Halzen, Phys. Rev. {\bf D 45} (1992) 377;
\bibitem{aspen} M. Block, E. Gregores, F. Halzen and G. Pancheri,
Phys.Rev.{\bf D60} (1999) 054024.
\bibitem{lundus}
R.~M.~Godbole and G.~Pancheri, hep-ph/9903331.
Proceedings of the {\it LUND workshop on photon interactions and
photon structure}, Aug. 1998, 217-227, Eds. T. Sjostrand and J. Jarsklog,
\bibitem{albert}
A.~De Roeck, Nucl.\ Phys.\ Proc.\ Suppl.\  {\bf 99A}, 144 (2001)
In the proceedings of {\it DIFFRACTION 2000: International Workshop
on Diffraction in High-energy and Nuclear Physics, Cetraro, Cosenza,
Italy, 2-7 Sep 2000.}
[hep-ph/0101076].
\bibitem{therise} 
D. Cline, F. Halzen and J. Luthe, {\it Phys. Rev. Lett.} {\bf 31}, 491 (1973).
G. Pancheri and C. Rubbia, {\it Nucl. Phys.} {\bf A 418}, 117c (1984).
T.Gaisser and F.Halzen, {\it Phys. Rev. Lett.} {\bf 54}, 1754  (1985).
G.Pancheri and Y.N.Srivastava, {\it Phys. Lett.} {\bf B 158}, 402 (1986).
\bibitem{eikminijets}
L. Durand and H. Pi, {\it Phys. Rev. Lett.} {\bf 58}, 58 (1987).
A. Capella, J. Kwiecinsky, J. Tran Thanh, {\it Phys. Rev. Lett.} {\bf 58},  
2015 (1987).
M.M. Block, F. Halzen, B. Margolis, {\it Phys. Rev.} {\bf  D 45}, 839 (1992).
A. Capella and J. Tran Thanh Van, {\it Z. Phys.} {\bf C 23}, 168 (1984).
P. l`Heureux, B. Margolis and P. Valin, {\it Phys. Rev.} {\bf D 32}, 1681 (1985).
 \bibitem{bn} A. Grau, G. Pancheri and Y. N. Srivastava, PR {\bf D60} (1999)
114020.
\bibitem{GRV} Gl\"uck, M., Reya, E., and Vogt, A., {\it Zeit. Physik C}\ {\bf
67}, 433 (1994).
\bibitem{collins}
J.C. Collins and G.A. Ladinsky, Phys. Rev. {\bf D 43} (1991) 2847.
\bibitem{GRVPH}
Gl\"uck, M., Reya, E., Vogt, A., {\it Phys. Rev. D}\ {\bf 46}, 1973 (1992). 
\bibitem{GRS} Gl\"uck, M., Reya, E., and Schienbein, I., {\it Phys.Rev. D}\
{\bf 60}, 054019 (1999); Erratum, {\it ibid.}\ {\bf 62}, 019902 (2000).
\bibitem{lcws00}
R.M.~Godbole and G.~Pancheri, {\bf hep-ph/0102188},
To appear in the Proceedings of {\it 5th International Linear Collider
Workshop (LCWS 2000), Fermilab, Batavia, Illinois, 24-28 Oct 2000. }
\bibitem{DL}Donnachie, A.,  and Landshoff, P.V., {\it Phys. Lett. B}\ {\bf
296}, 227 (1992), {\bf hep-ph/9209205}.
\bibitem{sjostrand}Schuler, G.A., and Sj\"ostrand, T., {\it Z.Phys.C}\ {\bf 73}, 677
(1997), {\bf hep-ph/9605240}. 
\bibitem{maor} Gotsman, E., Levin, E., Maor, U., and Naftali, E., {\it Eur.
Phys. J. C}\ {\bf 14}, 511 (2000), {\bf hep-ph/0001080}.
\bibitem{ttwu}Bourrelly, C., Soffer, J., and Wu, T.T., {\it Mod.Phys.Lett. A}\ {\bf
15}, 9 (2000). 
\bibitem{bkks}B. Badelek, B., M. Krawczyk, J. Kwiecinski  and A.M. Stasto,  
{\bf hep-ph/0001161}.
\bibitem{LCnote}R.M. Godbole, G. Pancheri and A. de Roeck, 
{\bf LC-TH-2001-030}.
\bibitem{yn1} A. Nakamura, G. Pancheri and Y. Srivastava, {\it Z.
Phys} {\bf C21} (1984) 243.
\bibitem{yn2} Y. Srivastava, S. Pacetti, G. Pancheri and A. Widom,
{\bf hep-ph/0106005.}
Proceedings of {Workshop on $e^+e^-$ interactions at low and medium
energies, SLAC, 2001}
\end{thebibliography}
\end{document}